\documentclass[11pt]{article}

\usepackage[utf8]{inputenc}
\usepackage[T1]{fontenc}
\usepackage{lmodern}
\usepackage[margin=1in]{geometry}
\usepackage{amsmath,amssymb}
\usepackage{graphicx}
\usepackage{xcolor}
\usepackage{microtype}
\usepackage{booktabs}
\usepackage{array}
\usepackage{tabularx}
\usepackage{caption}
\usepackage{flafter}
\usepackage{float}
\usepackage{etoolbox}

\AtBeginEnvironment{table}{\small}

\definecolor{resultcol}{HTML}{1F4E79}
\definecolor{passcol}{HTML}{1E7D32}
\definecolor{failcol}{HTML}{B00020}
\newcommand{\keyresult}[1]{\textbf{\textcolor{resultcol}{#1}}}
\newcommand{\passresult}[1]{\textbf{\textcolor{passcol}{#1}}}
\newcommand{\failresult}[1]{\textbf{\textcolor{failcol}{#1}}}
\usepackage[colorlinks=true,linkcolor=blue,citecolor=blue,urlcolor=blue]{hyperref}
\newcommand{\dataseturl}{https://huggingface.co/datasets/Setloop/What-Gradients-Add-to-Text-Leakage-in-Split-Language-Models-Counted-per-Token-and-per-Document}
\newcommand{\datasetrelease}{\dataseturl/tree/v1.0}

\title{What Gradients Add to Text Leakage in Split Language Models,\\ Counted per Token and per Document}
\author{Georgios Politis\\ {\small\texttt{geo.politis@gmail.com}} \and Setloop.io \and
  Evangelos Pappas\\ {\small\texttt{evanz.pappas@gmail.com}}}
\date{}

\begin{document}
\maketitle

\begin{abstract}
Split learning lets a client train a language model on a server without sending its text. The client runs the first layers itself and sends the server only their output, a vector of numbers for each token. During training, the server sends gradients back. We show that an observer at the split can rebuild most of the client's text from this traffic, and we measure how much the gradients help. On GPT-2, an attacker who holds only the publicly released weights of the client's layers recovers 94.20\% of tokens from the activations alone and 97.38\% when it also sees the gradients, 3.17 percentage points more (95\% interval [2.72, 3.64]). Counted by document, the difference is much larger. The attacker rebuilds 13.71\% of 32-token documents exactly without the gradients and 37.77\% with them, because a document only counts when every token is right. How we count also changes how good a defence looks. Secret mixup, which blends each outgoing vector with a decoy, stops the attacker from rebuilding almost any document exactly, yet the attacker still recovers 83--91\% of tokens. In a second experiment on GPT-2 and Qwen3-0.6B, where the server trains only a run of consecutive layers, the layer at which the run starts changes both model quality and leakage, even when the run's length is fixed. We recommend reporting leakage both per token and per document, and treating what a split model sends as being as sensitive as the text itself.
\end{abstract}

\section{Introduction}
\label{sec:intro}
Large language models are expensive to train and run, and split learning offers a compromise \cite{vepakomma2018split}. The client, which holds the private text, runs the first few layers of the model on its own hardware and sends their output to a server that runs the remaining layers. The raw text never leaves the client. What leaves instead is an \emph{activation} for every token, a vector of numbers that the client's layers computed from that token and the ones before it. During training, the server also sends something back, the \emph{gradient}, which tells the client's layers how to change to reduce the model's error. Both are computed from the private text, so the server, or anyone who can observe the traffic between the two, may be able to turn them back into text \cite{pasquini2021unleashing}.

\begin{figure}[H]
\centering
\includegraphics[width=\linewidth]{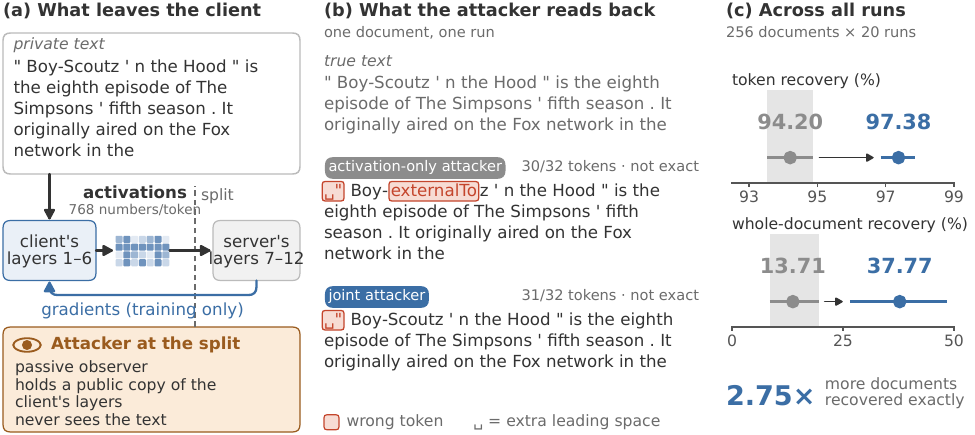}
\caption{\textbf{The gradients barely change how many tokens the attacker recovers, but nearly triple how many documents it rebuilds exactly.} (a) What leaves the client. The client runs GPT-2's first six layers and sends their activations to the server, which runs the other six. During training only, gradients come back. The attacker sits at the split, only watches this traffic, holds the publicly released GPT-2 weights and never sees the text. (b) What the attacker reads back from one document in one run: the true text, the reconstruction from activations only, and the reconstruction from activations and gradients, with wrong tokens highlighted. We chose this document as the one whose per-document accuracies are closest to the reported averages. The highlighted first token is a quotation mark predicted with a leading space, which looks the same in print. (c) Across all 256 evaluation documents and all runs: token recovery and whole-document recovery for both attackers, with 95\% intervals, and the 2.75-fold rise in documents rebuilt exactly.}
\label{fig:hero}
\end{figure}

Earlier work shows that each can be turned back into data on its own. Attacks on gradients recover training examples \cite{dlg,idlg}, including text \cite{lamp}, and attacks on a model's internal vectors recover the input that produced them \cite{vec2text,sipit}. The closest prior work, BiSR, attacks split training of large language models by optimising a guess of the text against both the activations and the gradients \cite{bisr2024}. In our own earlier study of one deployed split-training system, the returned gradient exposed which of the rows the client sent were decoys \cite{politis2026privacyfailure}. What these results do not tell a practitioner is how much the gradient adds when the activations alone already give most of the text away, and whether the answer depends on how success is counted.

We answer this with a controlled comparison. One attacker sees only the activations, while the other, which we call the \emph{joint attacker}, also sees the gradients. Everything else is held equal, including how many candidate tokens each may try at every position. Table~\ref{tab:signals} lists every signal an observer at the split can read and how we turn each one into a measured leak. We also ask a second, practical question. Some deployments hand only a run of consecutive layers to the server for training. For a run of fixed length, does it matter where in the model the run starts?

Figure~\ref{fig:hero} shows what we found for the first question. We split GPT-2, a twelve-layer model, after its sixth layer. The attacker gets the publicly released weights of those six layers, which anyone can download, rather than the client's trained copy. Working from the activations alone, it recovers 94.20\% of tokens. When it also sees the gradients, it recovers 97.38\%, 3.17 percentage points more, which clears the two-point bar we set before running the study. That looks like a small gain until we count whole documents. Without the gradients, the attacker rebuilds 13.71\% of the 32-token documents exactly. With them, it rebuilds 37.77\%. One wrong token is enough to spoil a document, so fixing the last few errors turns many near misses into exact copies. The gain is larger than the one we found in our earlier study \cite{politis2026privacyfailure}, which attacked a system with its defences switched on. Here the model has no defence, and the attacker already recovers most tokens from the activations.

Counting by document also changes how good a defence looks. Secret mixup, one of the two defences we test, blends each outgoing activation with a decoy taken from earlier training data. Under it, the attacker rebuilds at most 1.48\% of documents exactly, yet it still recovers 83--91\% of tokens. Anyone who judged this defence only by exact copies would think it much stronger than it is.

For the second question, we find that where the run starts matters, and not only its length. On GPT-2 and on Qwen3-0.6B, a 28-layer model, we train runs of two, four or six consecutive layers on the server, starting at one of the first four layers. Each run uses one of two defences, both of which compress what crosses the split: 8-bit rounding, or 4-bit rounding with gradient clipping and noise. We then measure what the split costs in model quality, which we call \emph{utility}: how much worse the split-trained model predicts held-out text than a model the client trains entirely by itself. We also measure leakage. Once we know how long the run is, its starting layer still explains differences in these outcomes in three of the four combinations for GPT-2 and in all four for Qwen3.

Our contributions are:
\begin{itemize}
\item \textbf{How to score defences.} We show that per-token and per-document scores can rank the same defence very differently, so a leakage report needs both (Section~\ref{sec:defences}).
\item \textbf{What gradients add.} We measure how much the returning gradients help an attacker that already reads the activations well, on documents that were never used to train the model or the attacker. On GPT-2, the gradients add 3.17 points of token recovery and nearly triple the share of documents rebuilt exactly. Noise or rounding applied to the gradients the attacker sees leaves most of this gain in place (Section~\ref{sec:joint}).
\item \textbf{Where offloading starts.} On two models, we show that when the server trains only part of the model, the layer where that part starts affects both model quality and leakage. Knowing the start helps predict the outcome of new training runs only under the 8-bit defence (Section~\ref{sec:p6}).
\item \textbf{Everything released.} We release the code, the exact inputs, the study plans we fixed in advance and the results of every run for both experiments (Appendix~\ref{app:evidence}).
\end{itemize}

A companion paper \cite{llmattacker2026framework} describes the attack framework used here, LLM-Attacker, and uses our main experiment as one of its case studies.

\section{Related Work}
\label{sec:related}
\paragraph{Split learning and its boundary.}
Split learning keeps the raw data with the client and shares only the activations at the layer where the model is split \cite{vepakomma2018split}. Keeping the data local does not by itself keep it private, because a malicious server can steer what the client's layers learn and then rebuild the client's private inputs \cite{pasquini2021unleashing}. Work on instance encodings asks the same question in general terms. An instance encoding transforms each private example before it is shared, so that a model can still be trained on it. One analysis asks whether such encodings can be private at all \cite{carlini2021instance}, and another bounds how well an encoding can be inverted \cite{maeng2022refil}. The activations at a split are an encoding of this kind, so our attacks measure how well one such encoding can be inverted in practice.

\paragraph{Rebuilding data from gradients and from internal vectors.}
Both signals that cross the split have been attacked before. Gradient inversion recovers training examples by starting from a dummy input and changing it until its gradients match the shared ones \cite{dlg,idlg}. LAMP applies this idea to text and uses a language model to keep its guesses fluent \cite{lamp}. Attacks on internal vectors work in a similar way. Vec2Text repeatedly corrects a guessed text until its embedding matches a target embedding \cite{vec2text}. SipIt relies on different inputs giving a language model different hidden states, so it can rebuild the input one position at a time, keeping the token whose hidden state matches the observed one \cite{sipit}. We run our own implementation of SipIt as one of our attacks. Rebuilding the input is not the only risk. Property inference, for example, learns attributes of the training data other than the data itself from shared model updates \cite{melis2019}. This is one reason we report more than whole-document recovery.

\paragraph{Leakage in split language models.}
The closest prior work is BiSR, which attacks split fine-tuning of large language models \cite{bisr2024}. It first guesses the text with a learned inversion model and then refines the guess by optimisation, so that it matches both the observed activations and the observed gradients. Our joint attacker uses the same two signals more simply. The gradient suggests candidate tokens, and the activations decide among them (Section~\ref{sec:background}). VFLAIR-LLM, a framework and benchmark for split learning of language models, includes attacks that recover the input \cite{vflair}, and our second experiment adapts its model-inversion attack. Our earlier study audited one deployed split-training system and found that the returned gradient exposed its decoy rows, so the system's privacy check had missed a leak \cite{politis2026privacyfailure}. This paper instead measures how much text an attacker recovers from undefended and defended models, how that depends on the unit of scoring, and whether it matters where offloaded layers start. Other work measures what a server sees during split inference \cite{fan2026serversee}, recovers both prompts and responses from split models \cite{gu2026pidi}, and protects the forward pass with differential privacy \cite{dpforward2023}. The companion study reports that its BiSR implementation agrees with a reference implementation at one split point. That agreement checks that the implementation is faithful but says nothing about how strong the attack is. The companion study also reports a separate comparison with an adaptation of BiSR, fixed in advance with its own budgets \cite{llmattacker2026framework}.

\section{Threat Model and Setup}
\label{sec:background}
\begin{figure}[H]
\centering
\includegraphics[width=\linewidth]{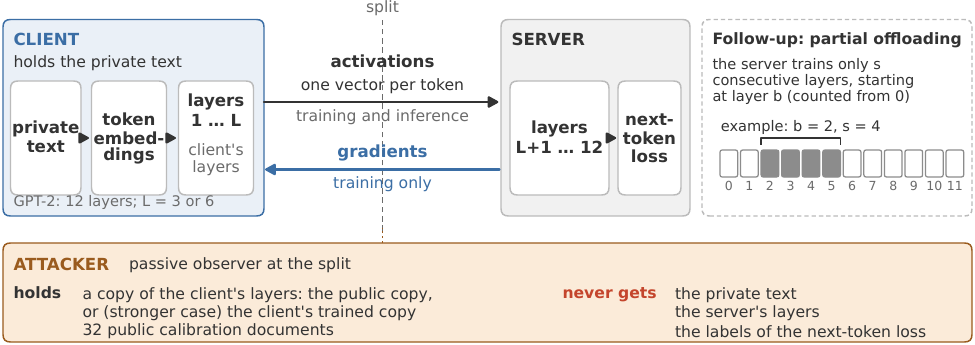}
\caption{\textbf{Split learning and what the attacker sees.} The client holds the private text and runs the first layers of the model, while the server runs the rest. In both training and inference, the client sends one activation per token across the split. In training only, the server sends back the gradient of the next-token loss with respect to those activations. The attacker watches this traffic without changing it. It holds a copy of the client's layers, either the publicly released weights or the client's trained copy, and a small set of public documents. It never gets the private text, the server's layers or the labels of the next-token loss. Inset: in the partial-offloading experiment, the server trains only a run of $s$ consecutive layers starting at layer $b$ (example: $b=2$, $s=4$).}
\label{fig:architecture}
\end{figure}

In both experiments the attacker is a passive observer at the split (Figure~\ref{fig:architecture}). It watches the traffic between the client and the server and changes nothing that either of them computes. We compare two such attackers. The \emph{activation-only attacker} sees the activations that the client sends. The \emph{joint attacker} also sees the gradients that the server sends back during training, which are the gradients of the model's next-token loss with respect to those activations. Both attackers probe the final trained model, whose weights no longer change. Neither collects gradients over the course of training.

Besides this traffic, the attacker holds a copy of the client's layers (the layers before the split) and their token embeddings, and we give it one of two copies. The \emph{public copy} is taken from the publicly released pretrained model at the split depth, and anyone can download it. The \emph{trained copy} is taken from the client's model after training, and we count the time needed to export it in the attack's cost. Our code that records the traffic uses the server's layers and the labels of the next-token loss, but the attacker never receives them. We use the evaluation text to score the attack and never give it to the attacker.

We score a reconstruction in three ways. \emph{Token recovery} is the share of tokens reconstructed correctly. Whole-document recovery, the strictest of the three, is the share of documents rebuilt with every token correct (Section~\ref{sec:intro}). \emph{Span recovery} is the share of annotated names, numbers and identifiers reconstructed exactly. We report all three because an attacker can rebuild few documents exactly and still recover most tokens.

\paragraph{Main experiment.}
The client's model is GPT-2 \cite{radford2019gpt2}, which has twelve layers and 124M parameters (\texttt{openai-community/gpt2}). We split it after its third or its sixth layer and write this split depth as $L \in \{3, 6\}$. Its inputs are 32-token passages, which we call documents, drawn from a pinned copy of WikiText-2 \cite{merity2016wikitext} (\texttt{Salesforce/wikitext}, \texttt{wikitext-2-raw-v1}). We divide the documents into three disjoint sets: 32 public calibration documents, which the attacker may use, 128 documents that train the client's model, and 256 evaluation documents, which the attacker must rebuild.

\subsection{How the attack works}
\label{sec:attack}
The attacker needs what crosses the split, a copy of the client's layers and a few public documents, and nothing else. Table~\ref{tab:signals} lists the signals it reads, when each one crosses, how the attack uses it and what it reveals. Here we explain how the attack turns the first two of these signals, the activations and the gradients, into text.

\begin{table}[H]
\centering
\small
\vspace{4pt}
\renewcommand{\arraystretch}{1.15}
\begin{tabular}{@{}>{\raggedright\arraybackslash}p{0.2\textwidth}>{\raggedright\arraybackslash}p{0.13\textwidth}>{\raggedright\arraybackslash}p{0.33\textwidth}>{\raggedright\arraybackslash}p{0.24\textwidth}@{}}
\toprule
Signal & When it crosses & How the attack uses it & What it reveals \\
\midrule
Activations, one vector per token & Training and inference & The ridge fit turns each activation into a short list of likely tokens, and a beam search keeps the guesses whose recomputed activations match (Figure~\ref{fig:pipeline}). SipIt also reads them. & Token, whole-document and span recovery (Section~\ref{sec:joint}) \\
\addlinespace
Gradients of the next-token loss & Training only & Each gradient carries a signal about the token that follows it, so it suggests half of the candidate tokens at each position. & What the gradients add to the same three scores \\
\addlinespace
Activations of the final trained model & Training & The ridge fit decodes each activation to the nearest token (adapted model inversion). & Tokens recovered beyond the majority control (Section~\ref{sec:p6}) \\
\addlinespace
Size of the returned gradient, per document & Training & The summed gradient size scores whether a document contains a given token. & Properties of the text, as ROC AUC (Section~\ref{sec:p6}) \\
\midrule
\emph{Controls:} another document's gradient; random tokens in place of the gradient's suggestions & Training & Same attack, with the gradient's information removed & Whether the gain comes from the gradient \\
\bottomrule
\end{tabular}
\caption{\textbf{The signals an observer at the split reads, and how each becomes a measured leak.} The first two rows belong to the main experiment and the next two to the partial-offloading experiment. The controls replace the gradient's information and show that the gain comes from what the gradient carries.}
\label{tab:signals}
\end{table}

\begin{figure}[H]
\centering
\includegraphics[width=\linewidth]{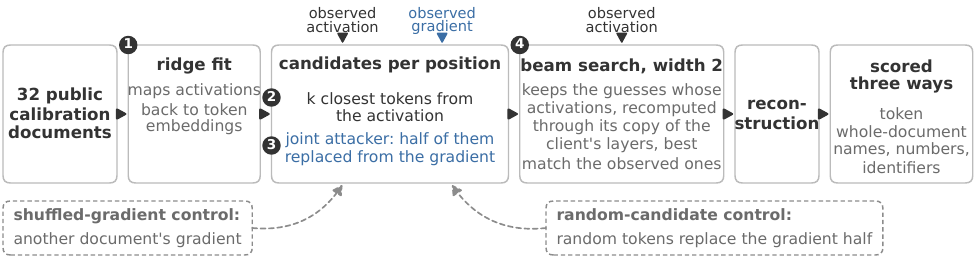}
\caption{\textbf{How the attacker rebuilds a document.} (1) From 32 public documents, the attacker learns a map from activations back to token embeddings. (2) At each position, it lists the $k$ most likely tokens. (3) The joint attacker replaces half of that list with tokens suggested by the gradient. The first position has no gradient help and keeps the list from the activations. (4) A beam search of width two builds the text from left to right, keeping the guesses whose activations, recomputed through the attacker's copy of the client's layers, best match the observed ones.}
\label{fig:pipeline}
\end{figure}

Both attackers rebuild a document in four steps (Figure~\ref{fig:pipeline}). They share the first, second and fourth steps and differ in the third.

First, the attacker learns a map from activations back to token embeddings, the vectors the model assigns to each token in its vocabulary. To do this, it passes the 32 public calibration documents through its copy of the client's layers, which gives it activations paired with the tokens that produced them. It then fits a ridge regression, a linear map kept stable by a penalty on its size, from each activation to the embedding of its token. We call this map the ridge fit.

Second, at each position of the observed document, the attacker applies the ridge fit to the observed activation and lists the $k$ tokens whose embeddings are closest to the result by cosine similarity. We call $k$ the candidate budget and use $k \in \{8, 16, 32\}$.

Third, the joint attacker replaces half of that list with tokens suggested by the gradient. The gradient comes from the next-token loss, whose target at each position is the document's next token, so the gradient at one position carries a signal about the token that follows it. The attacker keeps the top half of the list from the activations and adds the tokens whose embeddings best match this signal. Where the two halves overlap, it fills the gap with further tokens from the activation list. No earlier position predicts the first token, so the first position gets no gradient help and all its candidates come from the activations.

Fourth, a beam search of width two builds the text from left to right. At each position, it extends the two best partial texts with the candidate tokens, recomputes their activations through its copy of the client's layers, and keeps the two whose activations best match what it observed. Both attackers check the same number of partial texts, so what separates them is which candidates they try. The joint attacker uses the same two signals that BiSR combines \cite{bisr2024} but does not reproduce BiSR's optimisation. We compare it with the activation-only attacker, which has the same candidate budget and no gradient.

\paragraph{Other attacks and controls.}
Two controls remove what the gradient says about the text. The shuffled-gradient control gives the joint attacker the gradient of a different document. The random-candidate control keeps the activation half of the list and fills the other half with tokens drawn uniformly at random without replacement. A third attack, the backward-only heuristic, uses the gradient alone. Finally, we run SipIt \cite{sipit} with budgets of 32, 64 and 128 iterations per token. We implemented the published algorithm ourselves instead of using the authors' code. When the budget runs out, it returns the best candidate it has checked, so SipIt's guarantee of recovering the input exactly does not apply.

\paragraph{Size of the main experiment.}
In all, the main experiment runs 40 attack configurations, 28 without a defence and 12 with one, at both split depths and with both copies of the client's layers. We run each against ten trained models, one per victim seed, with two attack seeds, which gives 3{,}200 attack evaluations. Six configurations use activations only: the activation-only attacker at $k \in \{8, 16, 32\}$ and SipIt at 32, 64 and 128 iterations per token. Because inference sends activations only, we also count these six in the inference setting. This adds 480 evaluations, for 3{,}680 in total, each on the same 256 evaluation documents. These 480 evaluations reuse the training measurements against the same trained models instead of running the attacks again (Section~\ref{sec:activations}). Appendix~\ref{app:matrix} shows every configuration.

\paragraph{What we fixed in advance.}
We fixed the plan of the main experiment before running it. It includes the main test, its two-point pass mark and how we compute its interval, all described below. We fixed the plan of the second experiment before its new runs but after we knew the results of an earlier pilot, and we exclude the pilot from its evidence. We planned its Qwen3 replication separately, after a first run at 32-bit precision failed (Section~\ref{sec:qwen}). We annotated the name spans with AI assistance and fixed them before we knew any outcomes. Appendix~\ref{app:evidence} lists the released plans, inputs and execution records.

\paragraph{Statistics.}
Our main test compares the two attackers in one setting: depth $L{=}6$, the public copy of the client's layers and $k{=}32$. It subtracts the activation-only attacker's token recovery from the joint attacker's, and it passes if the lower end of its 95\% interval is above two percentage points. To build this interval, we first average over the two attack seeds. We then resample documents and trained models as two separate axes, keeping every document paired with every model. This procedure is called a crossed bootstrap (10 victim runs $\times$ 256 documents, 100{,}000 replicates, fixed seed). The interval therefore covers variation across documents and across training runs, with the training set held fixed.

We also test 47 further token comparisons, with a Bonferroni correction for testing many at once ($\alpha = 0.05/47$). They compare the joint attacker with the activation-only attacker, the two controls and the backward-only heuristic, across depths, copies of the client's layers and budgets. We compute every difference before rounding. For example, the main difference is $3.1714$ points to four decimals, so we report $3.17$ and not the difference of the rounded rates.

\section{Gradients Add Little to Token Recovery but Much to Whole-Document Recovery}
\label{sec:joint}
We now turn to the first question, what the gradients add to what the activations already reveal. Two of the results below have pass marks that we fixed before the study: the main test and the 47 further token comparisons. The rest are descriptive, which means we report them without a formal test. They cover whole-document and span recovery, the perturbed gradients and the defences. We planned no test of whether perturbing the gradient leaves recovery unchanged.

\subsection{Activations alone reveal almost every token}
\label{sec:activations}
We start with the activation-only attacker, because the gradients can only add to what the activations already reveal. Without any gradient, this attacker recovers $94.20$--$97.81\%$ of tokens at both split depths and with both copies of the client's layers (Table~\ref{tab:primary_by_knowledge}). Inference sends activations only, so these rates also describe an attacker that watches inference traffic.

\subsection{Gradients add three points of token recovery}
Against this high baseline, the main test passes. At depth six with the public copy of the client's layers, the joint attacker recovers $97.38\%$ of tokens and the activation-only attacker $94.20\%$. The paired difference is $+3.17$ percentage points, with a 95\% crossed-bootstrap interval of $[+2.72, +3.64]$ whose lower end clears the two-point mark. Of the tokens that the activation-only attacker gets wrong, the gradients correct about 55\%. The 47 further comparisons pass as well, each with the lower end of its Bonferroni-corrected interval ($\alpha = 0.05/47$) above zero.

Averaged over the two copies of the client's layers, the joint attacker has the highest token recovery at both depths (Table~\ref{tab:joint_reconstruction}), but it does not win in every condition. With the trained copy at depth three, SipIt slightly beats it ($98.89\%$ against $98.72\%$).

\begin{table}[H]
\centering
\begin{tabular}{lcc}
\toprule
\textbf{Attack} & \textbf{$L=3$} & \textbf{$L=6$} \\
\midrule
Backward-only heuristic & 24.0 & 50.0 \\
SipIt (128 steps) & 98.2 & 96.7 \\
Activation-only ($k{=}32$) & 97.6 & 94.5 \\
Joint, shuffled gradient ($k{=}32$) & 97.5 & 94.2 \\
\textbf{Joint, true gradient ($k{=}32$)} & \keyresult{98.5} & \keyresult{97.8} \\
\bottomrule
\end{tabular}
\caption{Token recovery (\%) of each attack at depths three and six, averaged over the public and the trained copy of the client's layers. The joint attacker has the highest token recovery at both depths. Each entry pools 10{,}240 document-level evaluations built from 256 distinct evaluation documents (10 victim seeds $\times$ 2 attack seeds $\times$ 2 copies of the client's layers). Values are rounded to one decimal. Table~\ref{tab:joint_reconstruction_full} gives two.}
\label{tab:joint_reconstruction}
\end{table}

\subsection{Whole-document recovery nearly triples}
\begin{figure}[H]
\centering
\includegraphics[width=\linewidth]{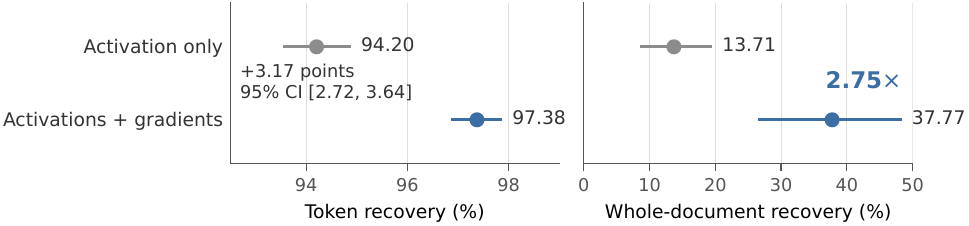}
\caption{\textbf{Gradients add little to token recovery but nearly triple whole-document recovery.} The panels show token recovery (left) and whole-document recovery (right) for the activation-only attacker and for the joint attacker, which also sees the gradients. GPT-2 is split after its sixth layer, the attacker holds the public copy of the client's layers, and the candidate budget is 32. Token recovery is the share of tokens reconstructed correctly. Whole-document recovery is the share of 32-token documents reconstructed without error. Bars are 95\% crossed-bootstrap intervals over documents and trained models. The left panel prints the paired difference in token recovery with its interval, and the right panel prints how many times higher whole-document recovery is with the gradients.}
\label{fig:teaser}
\end{figure}

Counted per document, the gradients help far more. A document needs all 32 of its tokens right to count as rebuilt, so a small gain on each token adds up over the document. At the setting of the main test, whole-document recovery rises from $13.71\%$ to $37.77\%$, a 2.75-fold increase (Table~\ref{tab:primary_by_knowledge} and Figure~\ref{fig:teaser}). The gain holds at both depths and with both copies, and it is largest with the trained copy at depth six, where recovery rises from $20.23\%$ to $58.55\%$.

The first token, which gets no help from the gradient, is often wrong. In the released predictions at the main setting, it is wrong in 2{,}408 of 5{,}120 activation-only reconstructions and 2{,}362 of 5{,}120 joint reconstructions. In about a third of these cases (820 and 806), the predicted token differs from the true one by nothing more than a leading space, such as a quotation mark or ``The'' written with a space before it, so the error is invisible in print. Because every token must be right, these first-token errors alone keep whole-document recovery below about 54\% for both attackers.

\begin{table}[H]
\centering
\begin{tabular}{llcccc}
\toprule
 & & \multicolumn{2}{c}{\textbf{Token recovery}} & \multicolumn{2}{c}{\textbf{Whole-document recovery}} \\
\cmidrule(lr){3-4} \cmidrule(lr){5-6}
\textbf{Depth} & \textbf{Copy} & \textbf{Activation-only} & \textbf{Joint} & \textbf{Activation-only} & \textbf{Joint} \\
\midrule
$L{=}3$ & public & 97.39 & 98.29 & 43.13 & 55.51 \\
$L{=}3$ & trained & 97.81 & 98.72 & 51.41 & 66.13 \\
\textbf{$L{=}6$} & \textbf{public} & \keyresult{94.20} & \keyresult{97.38} & \keyresult{13.71} & \keyresult{37.77} \\
$L{=}6$ & trained & 94.81 & 98.27 & 20.23 & 58.55 \\
\bottomrule
\end{tabular}
\caption{Token and whole-document recovery (\%) for the activation-only and joint attackers at $k{=}32$, by split depth and by the attacker's copy of the client's layers. The gradients' gain is small per token and large per document. Each entry pools 5{,}120 document-level evaluations over 256 distinct documents. The setting of the main test is in bold.}
\label{tab:primary_by_knowledge}
\end{table}

\subsection{Perturbing the observed gradient changes little}
\begin{figure}[H]
\centering
\includegraphics[width=\linewidth]{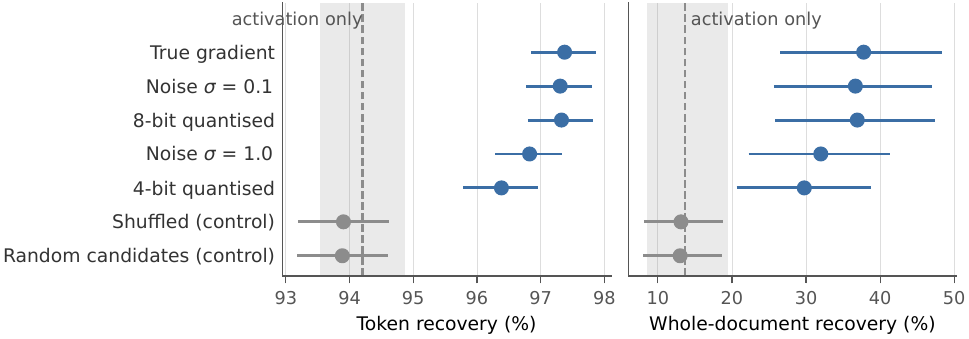}
\caption{Token recovery and whole-document recovery for each gradient condition at depth $L{=}6$, with the public copy of the client's layers and $k{=}32$ (5{,}120 document-level evaluations per condition over 256 distinct documents). Noise and rounding of the gradient leave most of its gain in place. The dashed line marks the activation-only attacker. Every condition stays above it except the shuffled-gradient and random-candidate controls, which remove the gradient's information and fall slightly below it.}
\label{fig:gradient_conditions}
\end{figure}

Because the gain comes from the gradients that the attacker observes, we next degrade what it observes. We either add Gaussian noise to the gradient or round it to fewer bits, a step called quantisation. The noise has standard deviation $\sigma$ times the root mean square (RMS) of each document's gradient. The rounding is symmetric and signed, and it is scaled by each document's largest absolute value. These perturbations change what the attacker sees and leave training untouched, so they are not deployed defences. Under them, token recovery stays close to its value with the true gradient (Figure~\ref{fig:gradient_conditions}). At depth six with the public copy, for example, the attacker recovers $97.31\%$ of tokens under noise of scale $\sigma = 0.1$ and $97.33\%$ under eight-bit rounding, against $97.38\%$ with the true gradient.

Stronger perturbations cost the attacker more but still leave most of the gain. It recovers $96.83\%$ of tokens under noise of scale $\sigma = 1.0$ and $96.38\%$ under four-bit rounding. By contrast, the two controls, in which the gradient contributes no information about the input, do worse than no gradient at all. The shuffled-gradient control recovers $93.91\%$ of tokens and the random-candidate control $93.89\%$, about 0.3 points below the activation-only attacker's $94.20\%$.

\subsection{Defences separate token recovery from whole-document recovery}
\label{sec:defences}
\begin{figure}[H]
\centering
\includegraphics[width=\linewidth]{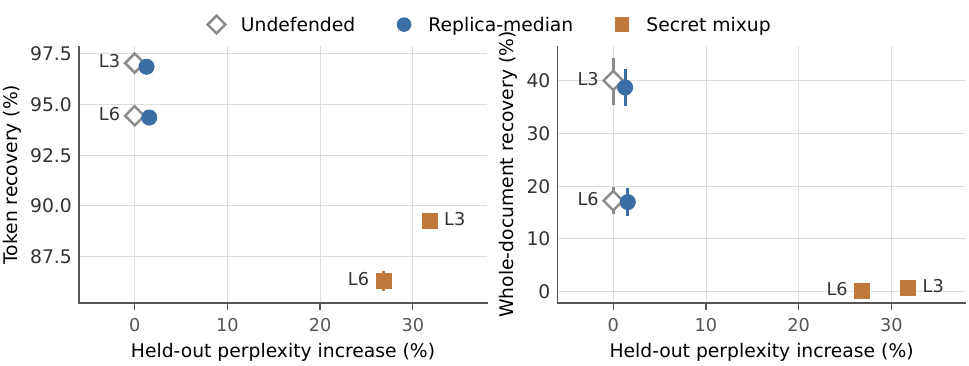}
\caption{Leakage under each defence, plotted against its cost in model quality. Leakage is token recovery (left) and whole-document recovery (right), pooled over the six attacks run against each defence at each depth. Cost is the rise in held-out perplexity over the final model trained without a defence. Secret mixup almost removes whole-document recovery at a $27$--$32\%$ perplexity cost, while $83$--$91\%$ of tokens stay recoverable. Replica-median changes token recovery little and lowers whole-document recovery by up to $7.7$ points at almost no cost in model quality.}
\label{fig:privacy_utility}
\end{figure}

The perturbations above leave training untouched. The two defences we test next change how the model is trained (Figure~\ref{fig:privacy_utility}), and we attack each with the same methods as the undefended model.

The first defence is \emph{secret mixup}. At each token position, it sends a weighted mix of the true activation and a decoy activation, with weight $\lambda = 0.75$ on the true one. The decoys are real activations recycled from earlier training data of the same run. A generator draws them without replacement for each chunk of training data sent, and its seed is derived with SHA-256 from the run's seed and the chunk's index. On the way back, the client runs the decoy alone through a parallel forward pass, which gives the decoy's gradient $g_{\rm decoy}$, and it corrects the returned gradient $g$ to $(g-(1-\lambda)\,g_{\rm decoy})/\lambda$. This correction is exact when the server's layers are linear and approximate otherwise. It also makes every returned gradient dense. Without it, decoy rows could receive gradients that are exactly zero, and our earlier study showed that anyone watching the traffic of a deployed system could spot such rows and discard them \cite{politis2026privacyfailure}. The second defence, \emph{replica-median}, protects the integrity of training rather than the privacy of the text. While the client's model is trained, it runs three independent copies of the server's layers in parallel, each with its own seed and a small seeded change to its weights, and combines their outputs by taking the median of each coordinate. The median stays correct as long as fewer than half of the copies are malicious. Under this defence, the attacker observes the activations sent to the copies.

Under secret mixup, the attacker still recovers $83.01$--$90.67\%$ of tokens across conditions, yet it rebuilds almost no documents exactly. Its best whole-document recovery is $1.48\%$, reached by the activation-only attacker at depth three with $k{=}8$ and either copy of the client's layers. With the public copy at $k{=}32$, the activation-only attacker rebuilds $0.31\%$ of documents at depth three and $0.04\%$ at depth six.

The annotated spans show which content secret mixup protects (Table~\ref{tab:span_classes}). Name recovery falls to $47.85$--$66.15\%$, from $67.36$--$97.44\%$ in the matching conditions without a defence. Numbers are protected far less, and the attacker still recovers $93.94$--$96.50\%$ of them and $90$--$100\%$ of identifiers. Names are also the largest annotated class, present in 249 of 256 documents, whereas the identifier class contains two spans. A document can contain spans of several classes.

\begin{table}[H]
\centering\small
\begin{tabular}{lrrrr}
\toprule
Class & Documents & Spans & No defence & Secret mixup \\
\midrule
\textbf{Name} & 249 & 716 & 67.36--97.44 & \keyresult{47.85--66.15} \\
Number & 142 & 297 & 97.64--100.00 & 93.94--96.50 \\
Identifier & 2 & 2 & 100.00--100.00 & 90.00--100.00 \\
\bottomrule
\end{tabular}
\caption{Exact span recovery (\%) by annotated class on the fixed evaluation set, without a defence and under secret mixup. Secret mixup protects names far more than numbers. Ranges are over the same 24 conditions: six attacks $\times$ two depths $\times$ two copies of the client's layers. Each rate pools exact span recoveries within an evaluation and then averages over its ten victim seeds and two attack seeds. Documents and spans are reused across conditions.}
\label{tab:span_classes}
\end{table}

A defence also costs model quality, and here the two defences differ sharply. We measure this cost as the change in held-out perplexity, which rises the worse the model predicts text it was not trained on. We take perplexity as a geometric mean and compare it with the final model trained without a defence. Secret mixup raises it by $31.8\%$ at depth three and $26.9\%$ at depth six. Replica-median, by contrast, costs almost nothing (perplexity $36.05 \to 36.51$--$36.62$) and leaves token recovery within two-thirds of a percentage point of the attacks without a defence. We measured model quality under a different release schedule from the attacks.

Although replica-median barely moves token recovery, it changes whole-document recovery by up to $7.7$ points. The largest change is for SipIt with the public copy at depth three, whose whole-document recovery falls from $43.77\%$ to $36.11\%$. Like secret mixup, then, replica-median leaves most tokens recoverable while it lowers whole-document recovery.

\section{Where the Server-Trained Layers Start Matters}
\label{sec:p6}
The main experiment fixes the split and asks what an attacker learns there. The second experiment asks a design question instead. Some deployments hand only a run of consecutive layers to the server for training. We call the number of layers in this run its \emph{length} $s$, and its first layer, counted from zero, its \emph{start} $b$. We ask whether, once the length is fixed, the start explains differences in model quality and leakage. We fixed the variance tests below and their pass marks before the new runs. The checks on held-out seeds are descriptive.

\paragraph{Design.}
The GPT-2 grid crosses lengths $s\in\{2,4,6\}$, starts $b\in\{0,1,2,3\}$, two defences and eight victim seeds, which gives 192 trained models. In addition, eight paired runs train the full model on the client alone, without protection. We call them all-local runs. All 200 runs completed and passed the evidence checks before analysis (Appendix~\ref{app:evidence}). We use the same 32-token document sets as in the main experiment.

Each run takes 128 AdamW \cite{loshchilov2019adamw} steps with batch size one and learning rate $5\times10^{-5}$, in one pass over the training documents in a seeded random order. The server trains the layers $[b,b+s)$, from layer $b$ up to but not including layer $b+s$. The attacker observes the split after $\max(1,b)$ layers, because the split software needs at least one layer on the client. A run that starts at layer 0 is therefore observed after layer 0, even though the server trains layer 0. For starts above one, moving the start moves the point where the attacker observes as well as the layers that are trained. Each run uses one of two defences, which act during training. The \emph{8-bit defence} rounds both the activations and the gradients to eight bits. The \emph{4-bit noisy defence} rounds them to four bits, clips the returned gradient at 1.0 and adds noise with standard deviation 0.1 times each document's gradient RMS, in the order clip, noise, round.

Utility measures what this split training costs in held-out loss. We define it as the extra held-out cross-entropy over the paired all-local model, $\Delta_U=\mathcal L_{\rm segment}-\mathcal L_{\rm all\ local}$, so a lower value means a smaller cost. For each seed, we also keep the held-out loss of the untouched pretrained model, as a control for what training the server's layers adds over the pretrained model.

\paragraph{Leakage.}
We measure leakage on the final trained model as well. We record the activations and the returned gradients after the defence, without updating the model, and the attacker uses the public copy of the client's layers, the 32 calibration documents and one fixed attack seed (17). The main attack adapts vanilla model inversion (VMI), the input-recovery attack of VFLAIR-LLM \cite{vflair}. Instead of the published optimisation, the attacker uses the ridge fit of our main attack (Section~\ref{sec:attack}). It fits the map on the 32 calibration documents and decodes each observed activation to the token whose embedding is nearest by cosine similarity.

A second attack targets properties of the text rather than the text itself. For each document, the attacker adds up the size (norm) of the returned gradient at each position. It uses this one score per document to guess three yes-or-no properties: whether the document contains the token with ID 286, 357 or 764. We picked these tokens because the share of calibration documents containing them is closest to one half, and broke ties by token ID. We use the true properties to score the attack and never give them to the attacker.

To fold both attacks into one number, we first score each. We compare VMI recovery with the majority control, which guesses the calibration documents' most frequent token at every position, and we score each property by the area under the ROC curve (AUC), which is 0.5 when the score carries no information. We then combine these scores into one \emph{normalised leakage index}
\begin{equation}
I=\max\left(0,\;a_{\rm VMI}-a_{\rm control},\;\max_{j=1,2,3}2|\mathrm{AUC}_j-0.5|\right).
\end{equation}
The absolute value makes an AUC below one half count as leakage just as one above it does. Because of the absolute values and the maximum, finite-sample noise alone can also make the index positive.

\paragraph{Analysis.}
For each defence and each outcome ($\Delta_U$ and $I$), we ask how much the start adds to a fit that already knows the length. To answer this, we compare a fit on length alone with a full fit that crosses length with start, over eight paired victim seeds. Partial $R^2$ is the share of the length-only fit's leftover variation, its residual sum of squares, that the full fit explains. It therefore captures the start and its interaction with length together.

A test passes if partial $R^2\geq0.10$ and a permutation test gives $p<0.05/4=0.0125$. The permutation test shuffles the start labels 10{,}000 times within each seed and length. We also report 98.75\% percentile intervals from 10{,}000 bootstrap draws over seeds. Because partial $R^2$ cannot be negative, we test for a zero effect with the permutation test rather than the interval.

\begin{table}[H]
\centering\small
\begin{tabular}{lllccc}
\toprule
Model & Defence & Outcome & Partial $R^2$ [98.75\% CI] & $p$ & Passes \\
\midrule
GPT-2 & 8-bit & Utility & \keyresult{1.00 [1.00, 1.00]} & \passresult{0.0001} & \passresult{Yes} \\
GPT-2 & 8-bit & Leakage index & \keyresult{0.99 [0.99, 1.00]} & \passresult{0.0001} & \passresult{Yes} \\
GPT-2 & 4-bit noisy & Utility & \keyresult{0.90 [0.85, 0.97]} & \passresult{0.0001} & \passresult{Yes} \\
GPT-2 & 4-bit noisy & Leakage index & \failresult{0.21 [0.13, 0.56]} & \failresult{0.019} & \failresult{No} \\
\midrule
Qwen3 & 8-bit & Utility & \keyresult{1.00 [1.00, 1.00]} & \passresult{0.0001} & \passresult{Yes} \\
Qwen3 & 8-bit & Leakage index & \keyresult{0.91 [0.85, 0.98]} & \passresult{0.0001} & \passresult{Yes} \\
Qwen3 & 4-bit noisy & Utility & \keyresult{0.92 [0.88, 0.98]} & \passresult{0.0001} & \passresult{Yes} \\
Qwen3 & 4-bit noisy & Leakage index & \keyresult{0.90 [0.88, 0.96]} & \passresult{0.0001} & \passresult{Yes} \\
\bottomrule
\end{tabular}
\caption{Variance tests for where the server-trained layers start, in GPT-2 and Qwen3. The start explains variation beyond length in seven of eight tests. The one failure is GPT-2 leakage under the 4-bit noisy defence. Each row gives the partial $R^2$ of the start and its interaction with length, a 98.75\% seed-bootstrap interval, and the permutation $p$ value from shuffling the start within each seed and length. A test passes if partial $R^2\geq0.10$ and $p<0.0125$. Values are rounded to two decimals. Appendix~\ref{app:p6} gives four decimals.}
\label{tab:placement}
\end{table}

\subsection{In GPT-2, the start matters in three of four tests}
In GPT-2, the start explains variation beyond length in utility under both defences and in leakage under the 8-bit defence (Table~\ref{tab:placement}). Under the 8-bit defence, it accounts for almost all leftover variation in both outcomes. Under the 4-bit noisy defence, the utility effect is also large, but the leakage effect reaches a partial $R^2$ of 0.21 and fails the corrected permutation test.

The means per placement show where these effects come from (Appendix~\ref{app:p6}). Under the 8-bit defence, runs that start at layer three pay several times the extra loss of earlier starts and leak a fraction as much. Under the 4-bit noisy defence, by contrast, VMI recovery stays near its control for every placement.

The variance tests use all eight seeds, so they cannot tell whether knowing the start also improves prediction for seeds that the fit has not seen. A separate check does this. We fit ridge models on seeds 11--66, evaluate them on seeds 77 and 88 without choosing a new division of seeds, and ask whether adding the start to a linear predictor based on length lowers the held-out root-mean-square error (RMSE) by at least the specified 10\%. Under the 8-bit defence it does, with RMSE falling from 1.5698 to 0.7594 for utility and from 0.3042 to 0.1337 for the leakage index. Under the 4-bit noisy defence, the falls are smaller, from 0.2154 to 0.2018 for utility and from 0.0783 to 0.0735 for leakage, and neither reaches 10\%. Finally, we report without a test the share of GPT-2's twelve layers that the server trains, $s/12$, as a stand-in for the share of forward and backward computation that is offloaded. We compute it from the model's shape instead of measuring it, and because all the layers have the same shape, it does not depend on the start.

\subsection{In Qwen3, the start passes all four tests but predicts held-out seeds only under the 8-bit defence}
\label{sec:qwen}
To see whether the GPT-2 pattern holds in a second model, we repeat the design on Qwen3-0.6B \cite{yang2025qwen3}, which has 28 layers. The replication keeps the same lengths, starts, defences and eight victim seeds, and each model is trained for 128 steps at learning rate $5\times10^{-5}$. Qwen's tokeniser cuts the same source documents into 32-token windows, so the token sequences differ between the two models. We choose the property tokens by the same rule on the calibration documents, now applied to Qwen tokens. Apart from the model, the tokeniser and these windows, the analysis differs from GPT-2's in one respect. The share of layers trained on the server has 28 layers as its denominator.

Qwen3 also needed more numerical precision, because a first run at 32-bit precision (FP32) stopped when the gradients became non-finite. We exclude the failed and partial runs from that attempt. The replication therefore uses 64-bit precision (FP64) for the model's weights, the backward computation, the normalisation layers (RMSNorm) and the attention softmax, and it keeps FP32 for the cross-entropy. All 192 trained models and eight all-local controls completed and were verified (Appendix~\ref{app:evidence}).

Unlike GPT-2, Qwen3 passes all four tests on both thresholds (Table~\ref{tab:placement}), with a partial $R^2$ of at least 0.90 and a permutation $p$ of 0.0001 in every case.

On held-out seeds, however, the two defences again differ. Under the 8-bit defence, adding the start lowers utility RMSE from 2.0948 to 1.3957 and leakage-index RMSE from 0.1812 to 0.1002, both beyond the 10\% mark. Under the 4-bit noisy defence, RMSE instead rises, from 2.2834 to 2.2947 for utility and from 0.2176 to 0.2256 for leakage. For Qwen3, the share of layers trained on the server is $s/28$, which again does not depend on the start.

\section{Discussion}
\label{sec:discussion}
The activations alone give most of the text away. The activation-only attacker recovers at least $94.20\%$ of tokens at every depth and with both copies of the client's layers, so the gradients add to traffic that already leaks. This is why we report both how much leaks without gradients and how much the gradients add. They add little per token and much per document, because whole-document recovery needs every token of the document to be right.

How we count recovery matters just as much when we judge a defence. Secret mixup cuts whole-document recovery to near zero and sharply lowers name recovery, yet it leaves $83.01$--$90.67\%$ of tokens and most numbers recoverable. Replica-median leaves a smaller gap of the same kind, barely changing token recovery but lowering whole-document recovery by up to $7.7$ points. So a defence can look strong by one score and weak by another.

In partial offloading, where the server-trained run starts matters as well as how long it is. Under the 8-bit defence, the start accounts for almost all leftover variation in both models and improves prediction on held-out seeds. Under the 4-bit noisy defence, the start still explains variation in utility in both models and in Qwen3 leakage, but it does not improve prediction on held-out seeds in either model.

Our result for the joint attacker also bears on BiSR \cite{bisr2024}, which combines the same two signals. Measured against a strong activation-only attacker with the same candidate budget, what the gradients are worth depends on whether recovery is counted per token or per document.

\section{Scope and Limitations}
\label{sec:limitations}
We studied two small models, GPT-2 with 124M parameters and Qwen3-0.6B, on 32-token documents from WikiText-2, with the split after the third or sixth layer. Larger models, longer documents and other kinds of text may leak more or less than we measured. We attacked each defence with the same methods we used on the undefended model. An attacker who knows the defence can always run these attacks, so our defended results show what such an attacker recovers at the least. Attacks designed around a defence can only add to it. In the second experiment we measured leakage with a ridge-based decoder and one attack seed, not with the published optimisation-based model inversion.

\section{Conclusion}
\label{sec:conclusion}
The activation-only attacker already reads the activations well. The joint attacker, which also sees the gradients at the split, recovers about three points more tokens and nearly triples whole-document recovery. On GPT-2 split after its sixth layer, with the public copy of the client's layers, token recovery rises from $94.20\%$ to $97.38\%$, and whole-document recovery rises from $13.71\%$ to $37.77\%$. Defences can pull these two scores apart. In partial offloading, the layer where the server-trained run starts affects model quality and leakage beyond what the run's length explains.

Practitioners should treat the activations sent during inference as sensitive, because the attacker recovered $94.20$--$97.81\%$ of tokens from them without any gradient. They should evaluate defences on token, whole-document and span recovery together, because secret mixup cut whole-document recovery to at most $1.48\%$ while $83.01$--$90.67\%$ of tokens remained recoverable. They should not expect noise or rounding of the observed gradient, at the strengths we tested, to remove what the gradient contributes, since token recovery stayed at $96.38$--$97.33\%$ under them. Finally, when offloading a run of layers, practitioners should choose and evaluate where it starts as well as its length, and check any placement rule on held-out seeds under the defence that will be deployed.

\appendix
\sloppy
\section{Evidence Availability and Reproduction}
\label{app:evidence}
We release what a reader needs to check and rerun the three experiments in this paper: the main experiment on GPT-2, the partial-offloading experiment on GPT-2, and its replication on Qwen3. For each experiment the release holds the code as it was run, the exact inputs, the plan (protocol) we fixed before the run, the per-run results and logs, and the analysis. The results, logs and code snapshots are in \href{\dataseturl}{the public Hugging Face dataset for this paper}, which also holds the final paper and its arXiv source. Tag \texttt{v1.0} freezes the complete release, and the root \nolinkurl{MANIFEST.sha256} covers every retained file except itself and \nolinkurl{.gitattributes}. The table at the end of this appendix lists the release paths and the experiment-specific checksums.

\paragraph{Getting and checking the files.}
The same three steps apply to every experiment. First, download the pinned release with \texttt{huggingface\_hub.snapshot\_download}, giving the dataset name, \texttt{repo\_type="dataset"} and \texttt{revision="v1.0"}. Second, check every file against the release manifest, from the downloaded folder:
\begin{verbatim}
shasum -a 256 -c MANIFEST.sha256
\end{verbatim}
Third, install the code snapshot in a fresh Python environment with \texttt{python -m pip install -e ./code}. The snapshot pins the research dependencies.

\paragraph{Main experiment on GPT-2.}
The main experiment has 3{,}680 evaluations, each one attack configuration run once on the 256 evaluation documents. Of these, 3{,}200 were executed and 480 reuse a measurement from another evaluation. We split the evaluations deterministically into 16 parts, called shards, across two GPU machines. The attacks took 336.1 hours of wall-clock time, summed over the executed evaluations. All 16 shards finished with no failed final evaluations. We replaced earlier aborted shard attempts, and the execution history records them and every restart. Each prediction transcript is written to a hash-chained evidence journal, in which every entry commits to the one before it, so a later edit is detectable. The release holds the plan fixed in advance (its protocol file), the input documents, the merged results, the table of contrasts behind the main comparisons, the 16 shard journals and the recorded binary captures. The same dataset revision also holds the historical audit of Appendix~\ref{app:history}. A verification guide in the release separates three levels of checking, covering file integrity, rescoring stored predictions, and regenerating predictions. To reproduce the headline numbers, assemble the shards as the guide describes and run the official analyser:
\begin{verbatim}
python3 experiments/analyze_validation.py --root <merged> \
  --output <dir> --protocol-version 3 \
  --expected-manifest-sha256 <digest>
\end{verbatim}
Here \texttt{<merged>} is the assembled results folder and \texttt{<digest>} is the checksum of its merged evidence journal. The analyser re-hashes every file, recomputes each evaluation's metrics from its stored predictions, and stops on any mismatch. We round each percentage independently to two decimal places. For example, $43.125\%$ becomes $43.13\%$.

\paragraph{Partial-offloading experiment on GPT-2.}
The partial-offloading experiment ran 200 training runs on one GPU in a random order fixed by a seed (listed in the table of release identifiers), using 1.369 hours of run time in total, summed over runs. The 200 runs exclude short infrastructure test runs. Every run sealed its evidence and pinned all captured files, and any failed or missing run blocks the analysis. The final-state evaluation keeps the changed victim-model parameters so that they can be replayed. The release has two parts under \nolinkurl{gpt2-partial-offloading/}. The \nolinkurl{raw-evidence/} folder contains the captures and checkpoints for all 200 verified runs, whose content hashes we verified remotely after upload. The \nolinkurl{reproduction-package/} folder contains the source snapshot, frozen protocol, pinned input, per-run index, analysis and reproduction instructions. The source snapshot is kept as a content-hashed archive. To reproduce the results, verify both internal manifests and follow the package instructions. These instructions keep the original file paths recorded inside the evidence, because rewriting them would change the hashed bytes.

\paragraph{Qwen3 replication.}
The replication repeats the 200-run comparison of the partial-offloading experiment on Qwen3. Its plan differs from the GPT-2 plan in the share of layers trained on the server and in one execution detail. Runs were sequential under a 24-hour GPU budget, with short test runs of both a server-trained and an all-local model. We stored captured tensors at 16-bit precision and checked them for non-finite values. After an earlier run at 32-bit precision (FP32) failed, diagnostic runs found the earliest stage that produced non-finite values, and two complete test runs had to pass before we admitted a fresh complete grid. Neither the diagnostic runs nor these test runs count among the 200 runs. The controller used 32{,}829.60 seconds. With the earlier failures and the test runs, the recorded time was 34{,}345.62 seconds (9.54 hours).

An independent analysis on a CPU machine then checked the recorded summary of each of the 200 runs and 1{,}400 evidence journals, using freshly downloaded scoring inputs and the complete archive inventory. All categorical and integer outputs and the normalised run summaries matched exactly. Three floating-point outputs differed by at most $5.56\times10^{-17}$, below the frozen tolerance of $10^{-10}$ (absolute plus relative). We hashed the full binary files on the machine that produced them when we archived them, and checked them through the inventory against the sealed manifest of each run, without downloading them again to the CPU machine.

The replication release has 77 files, and we verified it against its manifest, including the inventory and every file hash. It holds the source snapshots, the exact inputs and protocol, the primary and reproduced analyses, the environment records and the reproduction instructions. To reproduce the results, follow the same three steps and then these instructions. A separate qualification report records the analysis and inventory hashes. We keep it with the execution archive and do not publish it, because it contains internal execution paths. The released archive record and manifest are therefore the public record of the archived bytes.

\paragraph{Names used in the released files.}
The released files use internal labels, which the table below maps to the names in this paper. The companion study reuses the main experiment as its evidence base and adds its own separately frozen comparison with a BiSR adaptation.

\begin{table}[H]
\centering
\small
\begin{tabular}{@{}>{\raggedright\arraybackslash}p{0.31\linewidth}>{\raggedright\arraybackslash}p{0.6\linewidth}@{}}
\toprule
Label in the released files & Name in this paper \\
\midrule
protocol \texttt{v3}, v3 campaign & main experiment \\
corrected P6 follow-up & partial-offloading experiment (the second experiment) \\
original P6 pilot & the excluded pilot of the partial-offloading experiment \\
Qwen3 FP64 amendment & Qwen3 replication \\
$D_0$ & 8-bit defence \\
$D_1$ & 4-bit noisy defence \\
arm & attack configuration, an experimental condition, not one of the framework's registry modules \\
cell & one evaluation or trained run \\
cell anchor & recorded summary of one run \\
canary & short infrastructure test run \\
head; public head; exact head & the client's layers; their public copy; their trained copy \\
\texttt{SIPIT-128} & SipIt at 128 steps \\
grants & which copy of the client's layers the attacker holds \\
seam & split point \\
protocol v4 amendment & the companion study's comparison with a BiSR adaptation \\
P0--P5 & stages of an earlier programme, audited in Appendix~\ref{app:history} \\
\bottomrule
\end{tabular}
\caption{Internal labels in the released files and the names used in this paper.}
\end{table}

\noindent\begin{minipage}{\linewidth}
\paragraph{Release identifiers.}
The table below lists every release path and checksum referred to in this appendix. Each release item in the first column links to its folder under tag \texttt{v1.0}. The raw tier holds the partial-offloading experiment's captures and checkpoints, while the reproduction package holds its source, inputs, analysis and instructions.
\par\medskip
\begin{center}
\small
\def\UrlBreaks{\do\/\do\.\do\-\do\_\do\0\do\1\do\2\do\3\do\4\do\5\do\6\do\7\do\8\do\9\do\a\do\b\do\c\do\d\do\e\do\f\do\g\do\h\do\i\do\j\do\k\do\l\do\m\do\n\do\o\do\p\do\q\do\r\do\s\do\t\do\u\do\v\do\w\do\x\do\y\do\z}
\begin{tabular}{@{}>{\raggedright\arraybackslash}p{0.31\linewidth}>{\raggedright\arraybackslash}p{0.6\linewidth}@{}}
\toprule
Item & Identifier \\
\midrule
\multicolumn{2}{@{}l}{\emph{Main experiment on GPT-2 and historical audit}} \\
\href{\datasetrelease}{Dataset release} & \nolinkurl{v1.0} \\
Frozen protocol & \nolinkurl{protocols/validation_protocol.v3.json} \\
Inputs & \nolinkurl{corpora/validation-input.json} \\
Merged results & \nolinkurl{v3-campaign/merged/validation.json} \\
Contrasts & \nolinkurl{v3-campaign/review/contrasts.csv} \\
Verification guide & \nolinkurl{v3-campaign/EVIDENCE-RELEASE.md} \\
Historical audit manifest & \nolinkurl{MANIFEST-AUDIT-20260917.sha256} \\
\addlinespace
\multicolumn{2}{@{}l}{\emph{Partial-offloading experiment on GPT-2}} \\
\href{\datasetrelease/gpt2-partial-offloading/raw-evidence}{Raw evidence folder} & \nolinkurl{gpt2-partial-offloading/raw-evidence/} \\
Raw tier size & 11 files, 34,703,146,309 bytes \\
Raw tier manifest & \nolinkurl{gpt2-partial-offloading/raw-evidence/MANIFEST.sha256} \\
\href{\datasetrelease/gpt2-partial-offloading/reproduction-package}{Reproduction package} & \nolinkurl{gpt2-partial-offloading/reproduction-package/} \\
Protocol name & \nolinkurl{p6-corrected-followup-20260917-v2} \\
Run-order seed & 41778 \\
Protocol SHA-256 & \parbox[t]{\linewidth}{\raggedright\nolinkurl{88e2286339109d8c6356b75b5263a23647261136eb0d039ecca15075c03d533b}} \\
\addlinespace
\multicolumn{2}{@{}l}{\emph{Qwen3 replication}} \\
\href{\datasetrelease/qwen3-partial-offloading-replication/evidence}{Evidence folder} & \nolinkurl{qwen3-partial-offloading-replication/evidence/} \\
Archive record & \nolinkurl{qwen3-partial-offloading-replication/evidence/ARCHIVE.json} \\
Manifest & \nolinkurl{qwen3-partial-offloading-replication/evidence/MANIFEST.sha256} \\
\bottomrule
\end{tabular}
\end{center}
\end{minipage}

\clearpage
\section{Full Attack Matrix}
\label{app:matrix}
\begin{figure}[H]
\centering
\includegraphics[width=\linewidth,height=0.80\textheight,keepaspectratio]{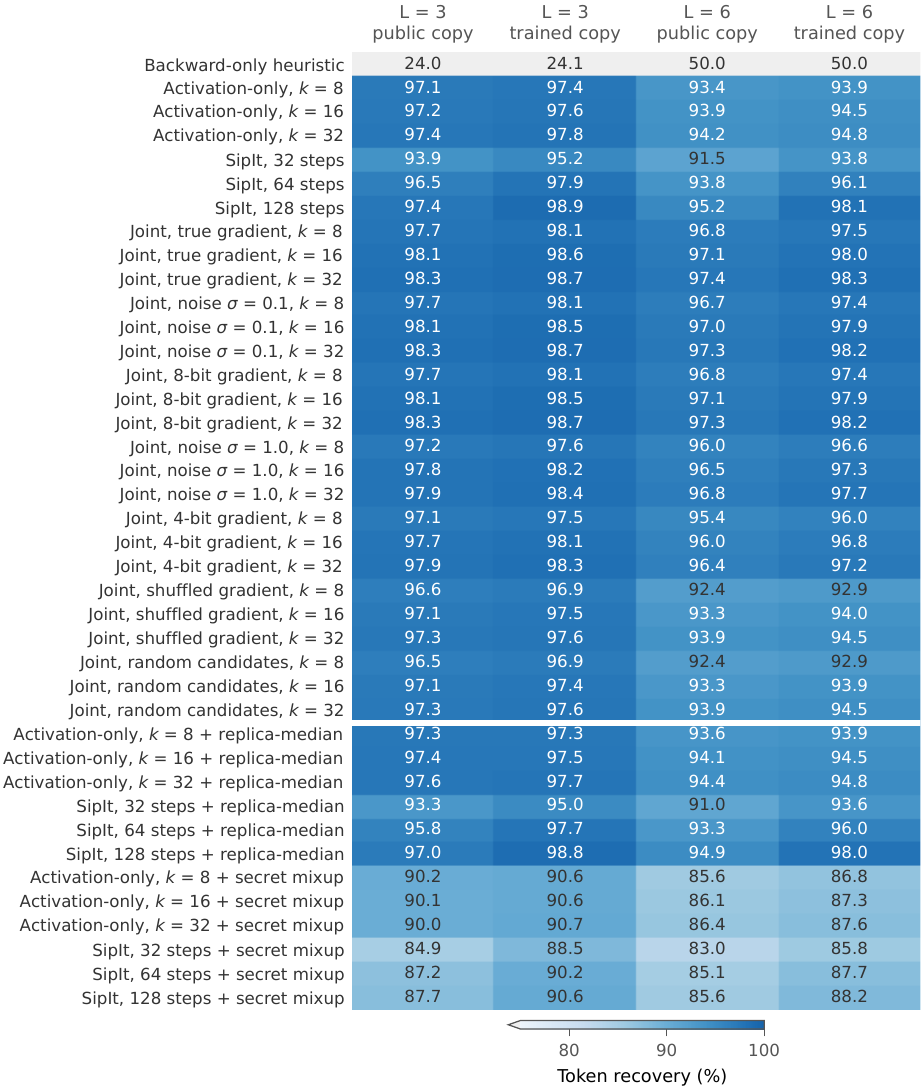}
\caption{Token recovery for all 40 attack configurations of the main experiment, at depths $L{\in}\{3,6\}$, with the public and trained copies of the client's layers. The 28 configurations without a defence are above the divider, the 12 defended ones below. Each entry aggregates 5{,}120 document-level evaluations over 256 distinct held-out documents. Depth three generally leaks more than depth six, and the trained copy leaks more than the public copy throughout. The backward-only heuristic reverses the depth ordering (${\approx}24\%$ at depth three against $50\%$ at depth six). In the defended rows, token recovery drops under secret mixup and stays close to the undefended rows under replica-median, as described in Section~\ref{sec:defences}.}
\label{fig:arm_heatmap}
\end{figure}

\begin{table}[H]
\centering
\begin{tabular}{lcc}
\toprule
\textbf{Attack} & \textbf{$L=3$} & \textbf{$L=6$} \\
\midrule
Backward-only heuristic & 24.04 & 50.01 \\
SipIt (128 steps) & 98.17 & 96.69 \\
Activation-only ($k{=}32$) & 97.60 & 94.51 \\
Joint, shuffled gradient ($k{=}32$) & 97.46 & 94.21 \\
\textbf{Joint, true gradient ($k{=}32$)} & \keyresult{98.50} & \keyresult{97.82} \\
\bottomrule
\end{tabular}
\caption{Token recovery (\%) from Table~\ref{tab:joint_reconstruction}, given to two decimals and averaged over the public and the trained copy of the client's layers. Each entry aggregates 10{,}240 document-level evaluations.}
\label{tab:joint_reconstruction_full}
\end{table}

\clearpage
\section{Partial-Offloading Experiment: Full-Precision Tests and Means per Placement}
\label{app:p6}
Tables~\ref{tab:p6} and~\ref{tab:qwen} give the variance tests of Table~\ref{tab:placement} to four decimals. In the released files and in the tables below, the 8-bit defence is $D_0$ and the 4-bit noisy defence is $D_1$.

\begin{table}[H]
\centering\small
\begin{tabular}{llcccc}
\toprule
Defence & Outcome & Partial $R^2$ & Interval & $p$ & Passes rule \\
\midrule
$D_0$ & Utility & \keyresult{0.9995} & [0.9993, 0.9998] & \passresult{0.00010} & \passresult{Yes} \\
$D_0$ & Leakage index & \keyresult{0.9898} & [0.9856, 0.9967] & \passresult{0.00010} & \passresult{Yes} \\
$D_1$ & Utility & \keyresult{0.8977} & [0.8493, 0.9673] & \passresult{0.00010} & \passresult{Yes} \\
$D_1$ & Leakage index & \failresult{0.2076} & [0.1283, 0.5571] & \failresult{0.01930} & \failresult{No} \\
\bottomrule
\end{tabular}
\caption{GPT-2 variance tests to four decimals, showing leftover variation explained by the start and its interaction with length, with 98.75\% seed-bootstrap intervals. The start passes three of four tests. A test passes when it meets both the partial $R^2$ threshold and the corrected permutation threshold.}
\label{tab:p6}
\end{table}

\begin{table}[H]
\centering\small
\begin{tabular}{llccc}
\toprule
Defence & Outcome & Partial $R^2$ & Interval & $p$ \\
\midrule
$D_0$ & Utility & \keyresult{0.9982} & [0.9972, 0.9994] & \passresult{0.00010} \\
$D_0$ & Leakage index & \keyresult{0.9089} & [0.8510, 0.9759] & \passresult{0.00010} \\
$D_1$ & Utility & \keyresult{0.9230} & [0.8833, 0.9815] & \passresult{0.00010} \\
$D_1$ & Leakage index & \keyresult{0.9040} & [0.8821, 0.9624] & \passresult{0.00010} \\
\bottomrule
\end{tabular}
\caption{Qwen3 variance tests to four decimals, reporting partial $R^2$ for the start and its interaction with length, with 98.75\% seed-bootstrap intervals and permutation $p$ values from shuffling the start within each seed and length. The start passes all four tests.}
\label{tab:qwen}
\end{table}

Table~\ref{tab:p6cells} reports GPT-2 means over eight victim seeds for every combination of length and start, measured on the final trained models. The release also keeps the value for each seed, all three property AUCs, per-document scores and crossed bootstrap intervals over trained models and documents for utility and VMI excess.

\begin{table}[H]
\centering\small
\begin{tabular}{rrrrrr}
\toprule
Defence & Length $s$ & Start $b$ & $\Delta_U$ & VMI excess & $I$ \\
\midrule
$D_0$ & 2 & 0 & 0.5342 & 0.9199 & 0.9199 \\
$D_0$ & 2 & 1 & 0.3854 & 0.9227 & 0.9227 \\
$D_0$ & 2 & 2 & 1.3715 & 0.6917 & 0.6917 \\
$D_0$ & 2 & 3 & 4.2361 & 0.0170 & 0.1468 \\
$D_0$ & 4 & 0 & 0.3173 & 0.9158 & 0.9158 \\
$D_0$ & 4 & 1 & 0.2342 & 0.9227 & 0.9227 \\
$D_0$ & 4 & 2 & 1.1799 & 0.6917 & 0.6917 \\
$D_0$ & 4 & 3 & 4.0686 & 0.0170 & 0.1766 \\
$D_0$ & 6 & 0 & 0.2377 & 0.9098 & 0.9098 \\
$D_0$ & 6 & 1 & 0.1345 & 0.9227 & 0.9227 \\
$D_0$ & 6 & 2 & 1.0823 & 0.6917 & 0.6917 \\
$D_0$ & 6 & 3 & 3.9813 & 0.0170 & 0.1854 \\
\midrule
$D_1$ & 2 & 0 & 5.1407 & 0.0011 & 0.1656 \\
$D_1$ & 2 & 1 & 4.4836 & -0.0247 & 0.2592 \\
$D_1$ & 2 & 2 & 4.6135 & -0.0282 & 0.1714 \\
$D_1$ & 2 & 3 & 4.4892 & -0.0282 & 0.1757 \\
$D_1$ & 4 & 0 & 4.5203 & -0.0035 & 0.1873 \\
$D_1$ & 4 & 1 & 4.2469 & -0.0247 & 0.2046 \\
$D_1$ & 4 & 2 & 4.5100 & -0.0282 & 0.1810 \\
$D_1$ & 4 & 3 & 4.3632 & -0.0282 & 0.1602 \\
$D_1$ & 6 & 0 & 4.4070 & -0.0127 & 0.2723 \\
$D_1$ & 6 & 1 & 4.1397 & -0.0247 & 0.2425 \\
$D_1$ & 6 & 2 & 4.4409 & -0.0282 & 0.1600 \\
$D_1$ & 6 & 3 & 4.3133 & -0.0282 & 0.1857 \\
\bottomrule
\end{tabular}
\caption{GPT-2 means across eight seeds for every placement. Utility is excess held-out loss in nats. VMI excess is recovery minus the majority control (fraction). $I$ is the normalised leakage index.}
\label{tab:p6cells}
\end{table}

\clearpage
\section{Audit of an Earlier Programme and Withdrawn Claims}
\label{app:history}
An earlier programme preceded the experiments in this paper. The released files label its stages P0 to P5. This appendix corrects the archival record. None of the claims in this paper rest on it. An audit of five internal execution hosts located 5{,}167 historical files and verified 3{,}720 manifest references with no mismatches. We did not recover another 772 references, chiefly tensors and corpus assets.

The owning project retired its pre-reset results on 10 September 2026, because the exact executed source was missing. The cited scaffold commit and dirty-tree hashes cannot reconstruct that source, so neither the recovered files nor recomputing their summaries reinstates the retired causal, corpus-frontier or P5 gate claims.

\paragraph{Qualification correction.}
P0's 30 checks were engineering tests that did not qualify the later privacy attacks, and the audit recovered no original \nolinkurl{PREREGISTRATION_P0.md}. The later qualification tested the recovered property attack of stage P4B, which used ridge regression on mean-pooled quantised latents to predict vowel fraction, punctuation fraction and newline count. It measured advantage as Pearson correlation minus the shuffled-null correlation, $r_{\rm attack}-r_{\rm null}$, rather than as a difference in $R^2$. Punctuation passed the recovered qualification rule, and the vowel and newline targets failed it. A separate structure attack predicted five next-byte classes with a two-layer MLP. These historical definitions differ from the token-presence AUCs of the partial-offloading experiment and cannot replace them.

\paragraph{Arithmetic correction for stage P2 (historical only).}
The recovered paired gain is frozen-random loss minus trained loss. It therefore measures what training adds and does not isolate the effect of training on the server. Table~\ref{tab:p2history} gives the eight raw-pair gains. Their mean is 0.016701851273 nats, their sample standard deviation is 0.003230494044, and their two-sided 95\% Student interval is [0.014001090665, 0.019402611881]. Against the 0.010-nat margin, the one-sided paired test gives $t(7)=5.867739630$, $p=0.000309678658$, and the margin-centred exact one-sided Wilcoxon test gives $W^+=35$, $p=0.0078125$. Against zero, the corresponding values are $t(7)=14.623140774$, $p=8.354586\times10^{-7}$ and $W^+=36$, exact $p=0.00390625$. Correcting the arithmetic does not repair the missing provenance.

\begin{table}[H]
\centering\small
\begin{tabular}{rrrr}
\toprule
Seed & Gain & Seed & Gain \\
\midrule
11 & 0.017826762050 & 55 & 0.020145211369 \\
22 & 0.018502995372 & 66 & 0.018460044637 \\
33 & 0.015580737963 & 77 & 0.018363717943 \\
44 & 0.014817791060 & 88 & 0.009917549789 \\
\bottomrule
\end{tabular}
\caption{Recovered P2 gains in nats, kept for the historical arithmetic audit.}
\label{tab:p2history}
\end{table}

The audit report and recovered archive are available under \nolinkurl{audit-20260917/} in the \texttt{v1.0} dataset release above. Newer external exploratory runs, labelled R1, are not a held-out confirmation, and this paper's claims do not use them.

\begingroup
\small
\bibliographystyle{plain}
\bibliography{main}
\endgroup
\end{document}